\documentclass[showpacs, pra,onecolumn,preprintnumbers ,amsmath, amssymb, superscriptaddress, aps]{revtex4-2}
\usepackage{color}
\usepackage{amsmath,amssymb}
\usepackage{pifont}
\usepackage{amssymb}  
\usepackage{bbold}
\usepackage{float}
\usepackage{subfloat}

\usepackage[caption=false]{subfig}
\usepackage{tikz}
\usepackage{makecell}
\usepackage{subfig}
\usepackage{pifont}   
\usepackage{graphicx} 
\graphicspath{{Figures/}}
\usepackage{dcolumn}  
\usepackage{bm}       
\usepackage{multirow} 
\usepackage{placeins}
\usepackage[colorlinks]{hyperref}
\usepackage{mathtools}
\usepackage{appendix}

\def \be{\begin{align}}
	\def \ee{\end{align}}
\def \bea{\begin{eqnarray}}
	\def \eea{\end{eqnarray}}

\begin{document}
	
	\title{Electron trapping in  graphene  quantum dots with magnetic  flux}
	\author{Mohammed El Azar}
	\affiliation{ Laboratory of Theoretical Physics, Faculty of Sciences, Choua\"ib Doukkali University, PO Box 20, 24000 El Jadida, Morocco}
	\author{Ahmed Bouhlal}
	\affiliation{ Laboratory of Theoretical Physics, Faculty of Sciences, Choua\"ib Doukkali University, PO Box 20, 24000 El Jadida, Morocco}
		\author{Abdulaziz D. Alhaidari}
	\affiliation{ Saudi Center for Theoretical Physics, P.O. Box 32741, Jeddah 21438, Saudi Arabia}
	\author{Ahmed Jellal}
	\affiliation{ Laboratory of Theoretical Physics, Faculty of Sciences, Choua\"ib Doukkali University, PO Box 20, 24000 El Jadida, Morocco}
	\affiliation{
		Canadian Quantum  Research Center,
		204-3002 32 Ave Vernon, BC V1T 2L7,  Canada}

	\begin{abstract}
		
It is known that the appearance of Klein tunneling in graphene makes it hard to keep or localize electrons in a graphene-based quantum dot (GQD). However, a magnetic field can be used to temporarily confine an electron that is traveling into a GQD. The electronic states investigated here are resonances with a finite trapping time, also referred to as quasi-bound states. By subjecting the GDQ to a magnetic flux, we study the scattering phenomenon and the Aharonov-Bohm effect on the lifetime of quasi-bound states existing in a GQD. We demonstrate that the trapping time increases with the magnetic flux sustaining the trapped states for a long time even after the flux is turned off. Furthermore, we discover that the probability density within the GQD is also clearly improved. We demonstrate that the trapping time of an electron inside a GQD can be successfully extended by adjusting the magnetic flux parameters.

	\end{abstract}
	\maketitle
	
	\section{Introduction}
	An arrangement of carbon atoms in a hexagonal shape, which forms covalent chemical bonds, makes up the two-dimensional material known as graphene. It is one of the most amazing materials due to its unique physical (mechanical, electrical and thermal) properties and the fact that charged particles behave as massless Dirac fermions at low energies \cite{novoselov2005two,neto2009electronic}. As a result,
	 electrons with typical nonrelativistic energy
	  may now be used to investigate relativistic effects. Since the creation of graphene, and thanks to its extraordinary electrical properties, many studies have been carried out into how it interacts with external fields \cite{goerbig2006electron,avetisyan2009electric}. Surprisingly, graphene exhibits a variety of special properties  showing that it can provide the ideal framework for the study of fundamental physics, such as the quantum Hall effect \cite{zhang2005experimental,jiang2007quantum,barlas2012quantum}, the Aharonov-Bohm effect \cite{recher2007aharonov,jackiw2009induced,schelter2012aharonov}, Landau quantization \cite{luican2011quantized,yin2017landau}, the Hofstadter butterfly spectrum \cite{nemec2007hofstadter}, etc.
	
Another exciting area of condensed matter physics is the prospect of localizing electrons in a given region of space despite the Klein tunneling phenomenon associated with massless spinors \cite{beenakker2008colloquium,allain2011klein}. Klein tunneling is a relativistic phenomenon involving full transmission of electrons over a potential barrier, whatever the height of the barrier, which explains the use of finite-dimensional circular quantum dots to trap electrons. However, in the case of GQDs subjected to specific favorable circumstances brought about by applying external electrostatic potentials, electron trapping has been documented for brief periods of time \cite{peres2006dirac,chen2007fock,silvestrov2007quantum,fehske2015electron}. Nonetheless, quasi-bound states are the electronic states of interest here. Such a state is often characterized by a finite lifetime (trapping time), in contrast to true bound states that live forever as in the case of an atom. For example, the brief confinement of the electron in the GQD is called quasi-localization, which disappears due to the Klein tunneling effect that causes electrons to leave the GQD. In addition, it has been demonstrated that the mass term \cite{jellal4358814electrons}, twisted light \cite{pena2022electron}, magnetic fields \cite{de2007magnetic,wang2009bound,pan2020quasi} or polarized light \cite{pena2022lifetime} can be utilized to create quasi-bound states in GQDs and prolong their duration (trapping time).
	
Based on these studies and in particular \cite{jellal4358814electrons,pena2022electrons}, we investigate the effect of a magnetic flux on the electron scattering phenomenon in a GQD placed in a uniform magnetic field and on how it affects the trapping time (lifetime) of quasi-bound states. This can be achieved by analyzing  the scattering efficiency $Q$, the probability density $\rho$ and the trapping time $\tau$ of electrons in GQDs. For that, we first determine the solutions of the Dirac equation and use the continuity condition at the interface to obtain the associated scattering efficiency outside and inside the GQD. Subsequently, we derive the trapping time from the imaginary part of the complex energy of the trapped electrons. Our numerical results show that an increase in the magnetic flux affects different physical quantities: (1) the scattering efficiency $Q$ reaches non-vanishing and significant values at zero magnetic field, (2) quasi-bound states start to be generated at low values of the GQD radius, (3) the probability of finding the electron inside the GQD is significantly improved, and (4) the trapping time of the quasi-bound states is significantly extended.

The structure of this paper is as follows. The exact solutions of the Dirac equation are derived in Sec. \textcolor{red}{II}  for an electron   passing through a magnetic GQD placed in a magnetic flux. The analysis of electron scattering in the present system is briefly presented in Sec. \textcolor{red}{III} where  we also specify the metrics needed to describe the scattering process. In Sec. \textcolor{red}{IV}, we present our numerical results based on the theory introduced in the previous sections. This numerical analysis highlights the main conclusions and provides clear justifications of our findings. Finally, we summarize our main results in the last section.
	
	\section{Theoretical model}\label{theory}
\begin{figure}[h]
		\centering 
		\includegraphics[scale=0.5]{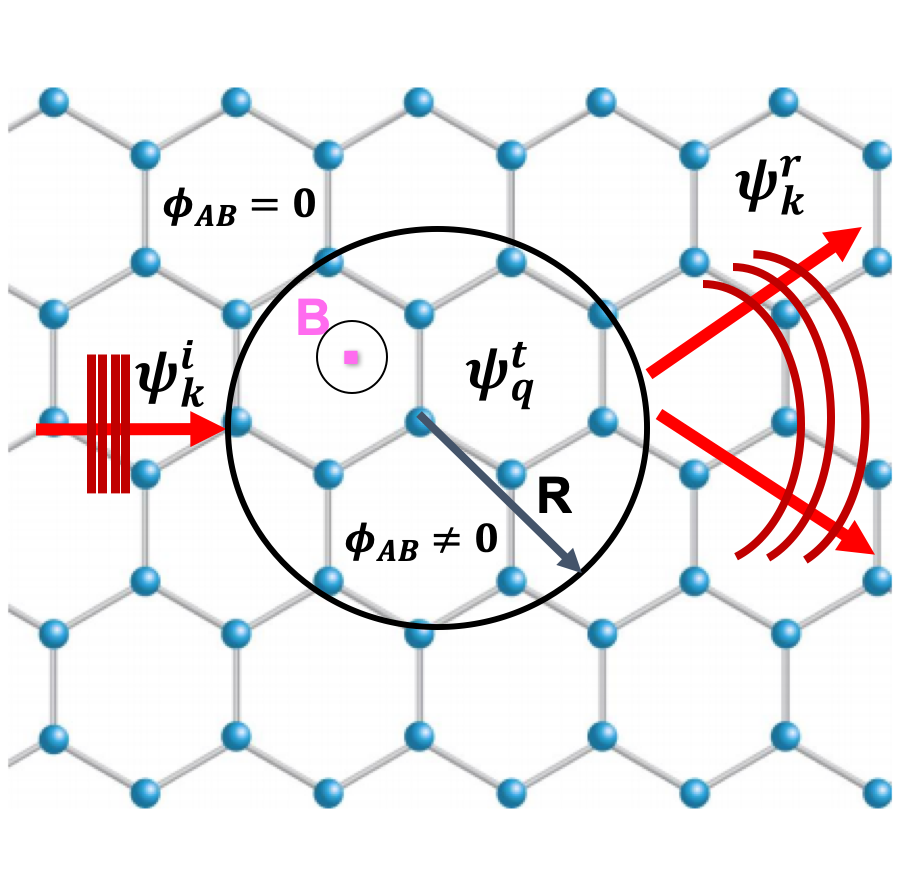}	
		\caption{(color online) A graphene quantum dot (GQD) of radius $R$ is confined by a constant magnetic field $B$ and exposed to a magnetic flux $\phi_{AB}$ in the $(x,y)$-plane. The plane wave  $\psi_k^i$ describes the state of the incident electron. Either an electron with energy $E $ is transmitted (wave function $\psi_q^t$) or reflected (wave function $\psi_k^r$).
%
	 }\label{figsystem}
	\end{figure}

	Let us consider a  graphene quantum dot (GQD) subjected to a a magnetic flux, which is made up of two regions  as depicted in Fig. \ref{figsystem}.   
We suggest the following single valley one-electron Hamiltonian to characterize the system: 
	\begin{equation}\label{Hamilt}
		H =v_F \vec \sigma \cdot (\vec p -e \vec A)
	\end{equation}
	where  $v_F = 10^6$ ms$^{-1}$  is the Fermi velocity, $(-e)$ is the  electron charge, and $\vec \sigma=(\sigma_x ,\sigma_y) $ are  the Pauli matrices. The vector potential $\vec{A}=\vec{A_1}+\vec{A_2}$ is chosen to be a linear combination of the symmetric gauge and the magnetic flux written in the polar coordinates $(r,\varphi)$ \cite{ikhdair2015nonrelativistic} 
\begin{align}
		\vec{A_1}=\frac{Br}{2}\vec{\varphi},\quad \vec{A_2}=\frac{\phi_{AB}}{2\pi r}\vec{\varphi}
\end{align}
where $\vec{\varphi}$ is unit vector. At this point, it is worthwhile emphasizing that outside the GQD, $\vec{A}$ does not have to vanish but could be a nonphysical gauge field of the form $\vec{A}=\vec{\nabla} F$ where $F$ is some scalar space-time function. 
Because of the cylinderical symmetry, we can write the Hamiltonian in
 polar coordinates knowing that $	\sigma_r =\cos \varphi \sigma_x +  \sin \varphi \sigma_y$ and  $	\sigma_r =-\sin \varphi \sigma_x +  \cos \varphi \sigma_y$. This yields
	\begin{equation}\label{ham3}
		H = \begin{pmatrix} 0 & -i\hbar v_F  e^{-i\varphi}\left[\partial_r - \frac{i}{r}\partial_\varphi - \frac{e Br}{2\hbar}-\frac{e\phi_{AB}}{2\pi\hbar r}\right]\\\\
		-i\hbar v_F  e^{i\varphi}\left[\partial_r + \frac{i}{r}\partial_\varphi + \frac{e Br}{2\hbar}+\frac{e\phi_{AB}}{2\pi\hbar r}\right]	&0 \\
\end{pmatrix}	
	\end{equation}		
	Since the Hamiltonian \eqref{ham3} commutes with the total angular momentum $J _z= -i\hbar \partial_ \theta +\frac{\hbar}{2}\sigma_z$, then we look for the eigenspinors that form a common basis to $H$ and $J_z$. They are
\begin{equation}\label{ansatz}
	\psi(r,\varphi)= e^{im\varphi}\dbinom{ \chi^A(r)}{ie^{i\varphi}\chi^B(r)}  
\end{equation}
where the azimuthal quantum number $m=0,\pm1,\pm2\cdots$. 
These will play a crucial role in analyzing the scattering phenomenon associated with the system.

To get the solutions of the problem, we solve the energy eigenvalue equation in each region. Indeed, for the  region outside the GQD where $B=0$ and $\phi_{AB}=0$, electrons scatter off the  GQD of radius $R$ in the absence of a magnetic flux. Consider an incident electron beam traveling in the $x$-direction under normal incidence with energy $E=\hbar v _F k$, where $k$ is the  wave number. 
Consequently, a plane wave may be used to describe the incident electron
\begin{align}
\psi_k^i(r, \varphi) =\frac{1}{\sqrt{2}} \sum_{m=-\infty}^{\infty} i^m e^{i m\varphi}\dbinom
	{J_m(k r)} 
	{i e^{i \varphi} J_{m+1}(k r)}
\end{align}
where $J_m(z)$ is the Bessel function of the first kind and the incident boundary conditions gives the upper component as $\frac{1}{\sqrt{2}} e^{i k x}=\frac{1}{\sqrt{2}} e^{i k \cos \varphi}$. Moreover, the scattering boundary conditions lead
to the following form of the reflected electron wavefunction that splits into partial waves 
 \cite{hewageegana2008electron,cserti2007caustics,heinisch2013mie,schulz2014electron}
\begin{equation}\label{e:26}
	\psi_k^r(r,\varphi) =\frac{1}{\sqrt{2}}\sum_{m=-\infty}^{\infty} a_m^r i^m \dbinom{H_m (kr) e^{im\varphi}}{iH_{m+1}(kr)e^{i(m+1)\varphi}}
\end{equation}
where $H_m(x) = J_m(x) + i Y_m(x)$ are the Hankel functions of the first kind as linear combinations of the Bessel functions$J_m$ and the Neumann functions $Y_m$, whereas $a_m^r$ could be determined by using the asymptotic behavior 
\begin{equation}\label{e:27}
	H_m(x) \underset{x\gg 1}{\sim} \sqrt{\frac{2}{\pi x}} e^{i(x-\frac{m\pi}{2}-\frac{\pi}{4})}.
\end{equation}

As far as the region inside the GQD that includes the magnetic field and the magnetic flux, one can obtain the transmitted solution starting with the Dirac wave equation 
$H \psi_q(r,\varphi)=E \psi_q(r,\varphi)$ to obtain 	
	\begin{subequations}\label{8}
		\begin{align}
&			\left(\partial_r -\frac{m-\mu}{r}+ \frac{r}{2 l_B^2}\right) \chi_q^A(r) =-q \chi_q^B(r) \label{8a}\\
		&	\left( \partial_r+\frac{m-\mu+1}{r} -\frac{r}{2l_B^2}\right) \chi_q^B(r) = q \chi_q^A(r) \label{8b}
		\end{align}
	\end{subequations}	
where we have set the magnetic length $l_B=(\hbar/e B)^{1/2}$, $\mu=\phi_{AB}/\phi_0$ and $\phi_0=h/e$. The wave number is linked to the energy  $E =
sv_F\hbar q$, where $s = +1$ stands for positive energy states
(conduction band) and, respectively, $s = -1$ for negative energy
states (valence band).
By injecting  \eqref{8a} into \eqref{8b}, we end up with  a second differential  equation for $\chi_q^A(r)$	
	\begin{equation}\label{e:9}
		\left( \partial_r^2+\frac{1}{r}\partial_r  + \frac{m-\mu+1}{l_B^2}-\frac{r^2}{4 l_B^4}-\frac{(m-\mu)^2}{r^2}+q^2\right)\chi_q^A(r) =0.
	\end{equation}
On the other hand, by injecting \eqref{8b} into \eqref{8a} we obtain the equation for $\chi_q^B(r)$ and conclude that the solution of $\chi_q^B(r)$ is simply obtained from $\chi_q^A(r)$ by the parameter maps
$m-\mu \longmapsto m-\mu+1$ and $q^2 \longmapsto q^2-\frac{2}{l_B^2}$.
Therefore, the energy gap for $\chi_q^A(r)$ is due to the mass term $\frac{\mu-m-1}{l_B^2}$, whereas the energy gap for $\chi_q^B(r)$ is due to the mass term $\frac{\mu-m}{l_B^2}$. In the limiting cases $r\longrightarrow 0$
and $r\longrightarrow \infty$, \eqref{e:9} can be written, respectively, as
\begin{align}
	& 	\left(  \partial_r^2+\frac{1}{r}\partial_r  - \frac{(m-\mu)^2}{r^2} \right) \chi_q^A(r)=0\\
	& 	\left( \rho \partial_\rho^2+\partial_\rho  - \rho \right) \chi_q^A(\rho)=0
\end{align}
showing that the solutions are proportional to $r^{\vert m-\mu\vert}e^{-\rho}$, with 
$\rho=\frac{r^2}{4 l_B^2} $. 
These can be used to propose the following ansatz:
	\begin{equation}\label{e:19}
		\chi_q^{A\pm} (r) =r^{\pm (m-\mu) }  e^{-r^2/4 l_B^2} \phi_q^{A\pm}(r)
	\end{equation}	
as general solution of \eqref{e:9} 
such that the "+" sign of the exponent is chosen for $m-\mu\ge 0$, while the "-" sign  is chosen for the opposite case $m-\mu<0$. Now, we perform the  variable change $ \eta=r^2/2 l_B^2 $,
which transforms \eqref{e:9} into the Kummer-type differential equations	 for $\phi_q^{A\pm}(\eta)$	
	\begin{subequations}\label{21}
		\begin{align}
			&\eta \partial_\eta^2 \phi_q^{A+}(\eta) +\left( m-\mu+1-\eta\right)  \partial_\eta \phi_q^{A+}(\eta) +\frac{l_B^2 q^2}{2} \phi_q^{A+}(\eta)=0  \label{21a}\\
			&\eta \partial_\eta^2 \phi_q^{A-}(\eta) +\left( 1-m+\mu-\eta\right)  \partial_\eta \phi_q^{A-}(\eta)
			+\left( m-\mu+\frac{l_B^2 q^2}{2}\right)  \phi_q^{A-}(\eta)=0. \label{21b} 
		\end{align}
	\end{subequations}
As a result, they have as solutions the confluent hypergeometric functions
	\begin{subequations}\label{18}
		\begin{align}
			&\phi_q^{A+}(\eta)=\prescript{}{1}{F}_1^{}\left(-\frac{l_B^2 q^2}{2},m-\mu+1,\eta\right) \label{18a}\\
			&\phi_q^{A-}(\eta)=\prescript{}{1}{F}_1^{}\left(-m+\mu-\frac{l_B^2 q^2}{2},1-m+\mu,\eta\right) \label{18b} 
		\end{align}
	\end{subequations}
Combining all the above results, we obtain the following solutions to the second-order differential equation \eqref{e:9}
		\begin{subequations}\label{22}
		\begin{align}
			& \chi_q^{A+}(\eta)= \eta^{\vert m-\mu\vert / 2} e^{-\eta / 2}\prescript{}{1} F_1\left(-\frac{l_B^2 q^2}{2}, m-\mu+1, \eta\right)
			\\
			&
			\chi_q^ {A-}(\eta)= \eta^{\vert m-\mu\vert / 2} e^{-\eta / 2} \prescript{}{1}F_1 \left(-m+\mu-\frac{l_B^2 q^2}{2},1-m+\mu, \eta\right)
		\end{align}	
	\end{subequations}	
		As noted below \eqref{e:9} the second spinor component $\chi_q^{B \pm}(r)$ is obtained from the first $\chi_q^{A \pm}(r)$ by the parameter map $m-\mu \longmapsto m-\mu+1$ and $q^2 \longmapsto q^2-\frac{2}{l_B^2}$. However, the parameter map does not determine the overall normalization that could be obtained using the differential relations \eqref{8a} or \eqref{8b}. As a result, we get
		\begin{subequations}
		\begin{align}
		& \chi_q^{B+}(\eta)=\frac{q l_B / \sqrt{2}}{\vert m-\mu\vert+1} \eta^{(\vert m-\mu\vert+1) / 2} e^{-\eta / 2}\prescript{}{1} F_1\left(1-\frac{l_B^2 q^2}{2}, m-\mu+2, \eta\right)
		\\
		&
		\chi_q^ {B-}(\eta)=-\frac{\vert m-\mu\vert}{q l_B / \sqrt{2}} \eta^{(\vert m-\mu\vert -1)/ 2} e^{-\eta / 2} \prescript{}{1}F_1 \left(-m+\mu-\frac{l_B^2 q^2}{2},-m+\mu, \eta\right)
	\end{align}	
	\end{subequations}	
	Finally, the solution inside the GQD can be obtained from the above analysis as
	\begin{equation}\label{e:28}
		\psi_q^t(r,\varphi) =\sum_{m=-\infty}^{\infty} a_m^t  \dbinom{\chi_q^{A\pm} (r) e^{im\varphi}}{i \chi_q^{B\pm}(r)e^{i(m+1)\varphi}}
	\end{equation}
	where the coefficients $ a_m^t $ are to be determined by the boundary conditions at $r=R$.
	Next, we will show how these results can be used to study the scattering problem and related matters.

	\section{Scattering phenomenon}

	To study the scattering problem of the system, we  need to determine the scattering coefficients $a_m^r$ and $a_m^t$ using the continuity of eigenspinors at the boundary condition $r=R$ 	
	\begin{equation}\label{e:29}
		\psi_k^i(R,\varphi) +\psi_k^r(R,\varphi) =\psi_q^t(R,\varphi). 
	\end{equation}
	After simplification, we end up with  two equations of $a_m^r$ and $a_m^t$ 
	\begin{subequations}\label{30}
		\begin{align}
		&\frac{1}{\sqrt{2}} i^m J_m(kR)+\frac{1}{\sqrt{2}} i^m a_m^r H_m(kR) =a_m^t \chi_q^{A\pm} (R) \label{30a}\\
		&	\frac{1}{\sqrt{2}} i^{m+1} J_{m+1}(kR)+\frac{1}{\sqrt{2}} i^{m+1} a_m^r H_{m+1}(kR) = i a_m^t \chi_q^{B\pm} (R)  \label{30b} 
		\end{align}
	\end{subequations}
	and they  can be solved to obtain  
	\begin{subequations}\label{31}
		\begin{align}
			a_m^{t\pm}(\mu)&=\frac{i^m} {\sqrt{2}}\  \frac{J_m(kR) H_{m+1}(kR)- J_{m+1}(kR) H_{m}(kR)}{
				H_{m+1}(kR)\chi_q^{A\pm} (R)	-H_{m}(kR)
					\chi_q^{B\pm} (R)} \label{31a}\\
			a_m^{r\pm}(\mu)&=\frac{-J_m(kR)\chi_q^{B\pm} (R)+J_{m
					+1}(kR)\chi_q^{A\pm} (R)}{H_m(kR)\chi_q^{B\pm} (R)-H_{m
					+1}(kR)\chi_q^{A\pm} (R) }  \label{31b} 
		\end{align}
	\end{subequations}
	which are  magnetic flux $\mu=\phi_{AB}/\phi_0$-dependent. Note that,  
the following identity is used to simplify \eqref{31a} for 	$a_m^{t\pm}(\mu)$
\begin{equation}
	J_{m+1}(z) H_m(z)- J_m(z) H_{m+1}(z)=\frac{2i}{\pi z}.
\end{equation}
	
		We close  by defining the main quantities used to describe the diffusion process. We consider the density of states 
	\begin{equation}
	\rho= \psi^\dag \psi
	\end{equation}
	and the current density
	\begin{equation}
	\vec j= \psi^\dag \vec \sigma \psi
	\end{equation}
where the function $\psi$  depends  on the region with $\psi=\psi_q^t$  inside the GQD and $\psi=\psi_k^i+\psi_k^r$  outside. The diffusion efficiency is calculated by dividing the diffusion cross-section by the geometric cross-section \cite{belokda2022electron,heinisch2013mie,schulz2014electron}
	\begin{equation}\label{e:33}
Q=\frac{\sigma}{2 R}=\frac{4}{k R}\sum_{m=-\infty}^{+\infty} \vert a_m^r(\mu)\vert^2.
\end{equation}
In order to estimate the trapping time (lifetime) of the quasi-bound states  inside the GQD, we make another analysis in terms of the complex incident energy  \cite{narimanov1999semiclassical}
\begin{align}
	E=E_r-iE_i
\end{align}
where $E_r$ represents the resonance energy and $E_i$ gives the trapping time of the quasi-bound state $\tau$, with consideration for the unique graphene linear dispersion law $\tau=\frac{\hbar}{E_i}$. As a result, the wave number $k$ also becomes complex
 \begin{align}
	k=k_r-ik_i
\end{align}
 and the trapping time is defined by
\begin{align}
\tau=\frac{1}{v_F k_i}.
\end{align}
 To perform this analysis, we use the continuity to determine the complex energy for the incidence  by finding the complex poles of the transmission and reflection coefficients \eqref{31}. As the kinetic energy of the incident electron is unaffected by the magnetic field and magnetic flux, we assume $q = k$ and deal with the following transcendental equation for $k$ \cite{hewageegana2008electron}:
	\begin{equation}\label{e:34}
\frac{\chi_k^{A\pm} (R)}{\chi_k^{B\pm} (R)}=\frac{H_m(kR)}{H_{m+1}(kR)}.
\end{equation} 
The above results will be numerically analyzed to identify the main features of the system. We plot all scattering quantities under various conditions and compare to results in the literature.

	\section{RESULTS AND DISCUSSIONS}\label{res}

\begin{figure}[ht]
 	\centering
 	\includegraphics[scale=0.4]{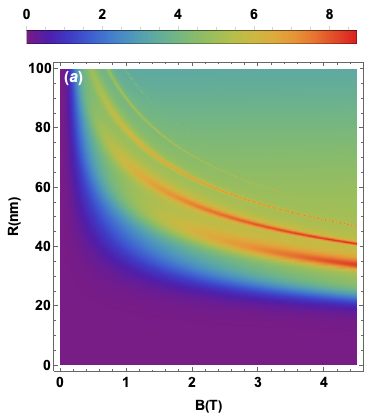} \includegraphics[scale=0.4]{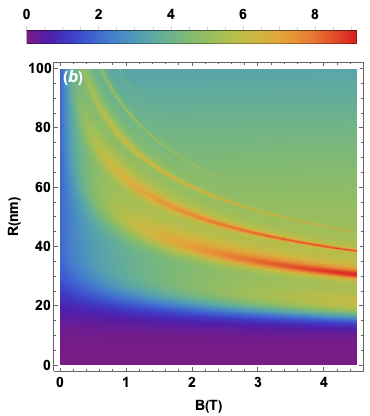}
 	\includegraphics[scale=0.4]{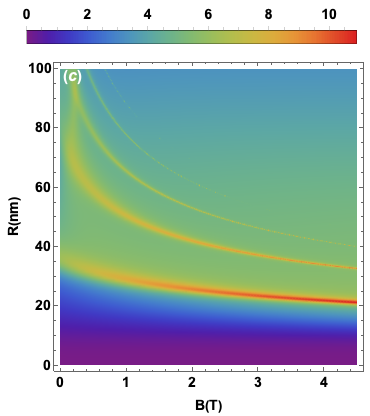} \includegraphics[scale=0.6]{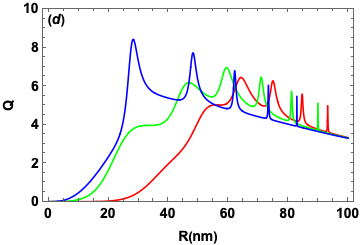} \includegraphics[scale=0.6]{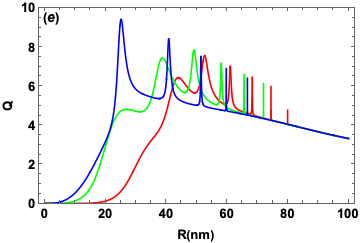}
 	\caption{(color online) (a,b,c): Scattering efficiency $Q$ is plotted versus the radius $R$ and the magnetic field $B$ for an incident energy $E=20$ meV and different values of magnetic flux  (a): $\mu=0$, (b): $\mu=\frac{1}{2}$, (c): $\mu=\frac{3}{2}$. (d,e): Scattering efficiency $Q$ as a function of  $R$ for $E=20$ meV,   $\mu=0$ (red color), $\mu=\frac{1}{2}$ (green color), $\mu=\frac{3}{2}$ (blue color), and   (d): $B=1.2$ T, (e): $B=2.2$ T.}
 	\label{fig2art}
 \end{figure}

We present the main numerical results to describe the electron scattering phenomenon in a practical way. That is why we make the analysis in terms of scattering modes, each mode corresponds to a number of angular momentum $m$ indexed by $m\in \mathbb{Z}$. We will concentrate on the modes involved in the scattering process and neglect the others. In Fig. \ref{fig2art}, the incident energy $E=20$ meV. In Figs. \ref{fig2art}(a,b,c), we plot the scattering efficiency $Q$ as a function of the magnetic field intensity $B$ and the radius $R$ of the GQD for the following AB-flux field values $\mu=0,1/2,3/2$, and in Figs. \ref{fig2art}(d,e) where $B$ takes the values 1.2 T and 2.2 T respectively, we plot $Q$ as a function of the radius $R$ for the same previous magnetic flux values $\mu$. In Fig. \ref{fig2art}a, we can clearly see that in the absence of magnetic flux, the interaction is very weak in a radius range from 0 to 32 nm. This radius range is reduced when we increase the magnetic flux, as shown in Figs. \ref{fig2art}(b, c), and above this radius range, wide and narrow bands start to appear. The notable increase in scattering efficiency $Q$ is due to the excitations of the specific scattering modes, each mode corresponds to a state of the angular momentum number $m$. In Figs. \ref{fig2art}(d,e), we see that as the magnetic flux  increases, the most relevant resonance peaks are shifted to smaller values of the GQD radius $R$. We also see that the scattering efficiency is improved with increasing magnetic flux and takes 8.4 as the maximum value at $B=1.2$ T and $\mu=3/2$ as shown in Fig. \ref{fig2art}d, until it reaches value 9.5 at $B=2.2$ T and $\mu=3/2$ as shown in Fig. \ref{fig2art}e.

\begin{figure}[H]
	\centering
	\includegraphics[scale=0.38]{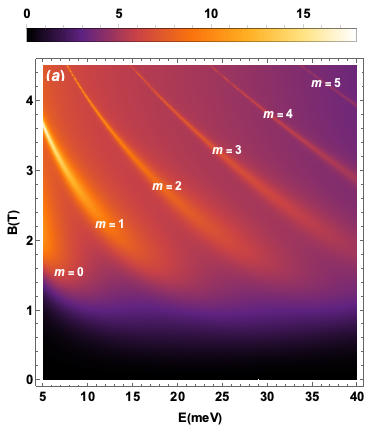}\ \ \ \ \ \ \includegraphics[scale=0.38]{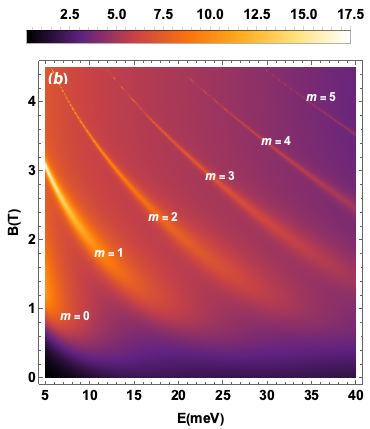}\ \ \ \ \ \
	\includegraphics[scale=0.38]{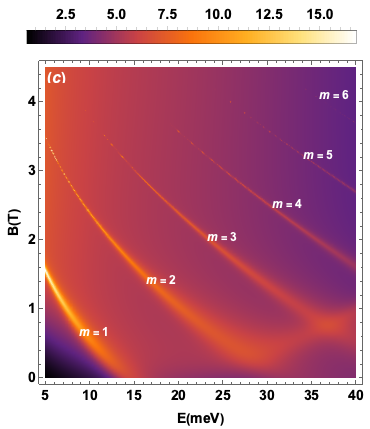}
		\caption{(color online) Scattering efficiency $Q$ as a function of magnetic field $B$ and incident energy $E$ for $R=50$ nm and different values of magnetic flux (a): $\mu=0$, (b): $\mu=\frac{1}{2}$, (c): $\mu=\frac{3}{2}$.}
		\label{fig3art}
	\end{figure}

 In Fig. \ref{fig3art}, we plot the scattering efficiency $Q$ as a function of the incident energy $E$ and the magnetic field $B$ for a GQD of radius $R=50$ nm and three values of magnetic flux  (a): $\mu=0$, (b) :$\mu= 1/2$ and (c) :$\mu=3/2$. In Fig. \ref{fig3art}a with $\mu=0$, we observe six somewhat oscillating bands of very large values of $Q$, which correspond to the number of angular momentum $m=0,1,2,3,4,5$. It is clearly seen that there is no interaction below the magnetic field value $B\approx1$ T. In Fig. \ref{fig3art}b, one always sees the appearance of six bands, but the interaction inside the GQD starts with smaller values of $B$ compared to the results exhibited in Fig. \ref{fig3art}a. Fig. \ref{fig3art}c shows the suppression of the band corresponding to the $m=0$ scattering mode with the appearance of a narrow band corresponding to the $m=6$ scattering mode. The most important point to note here is that the interaction is significant even in the absence of a magnetic field ($B=0$ T), as displayed in Fig. \ref{fig3art}c.
 
\begin{figure}[ht]
		\centering
		\includegraphics[scale=0.47]{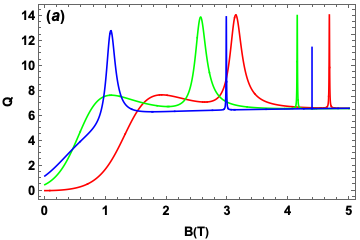}\includegraphics[scale=0.47]{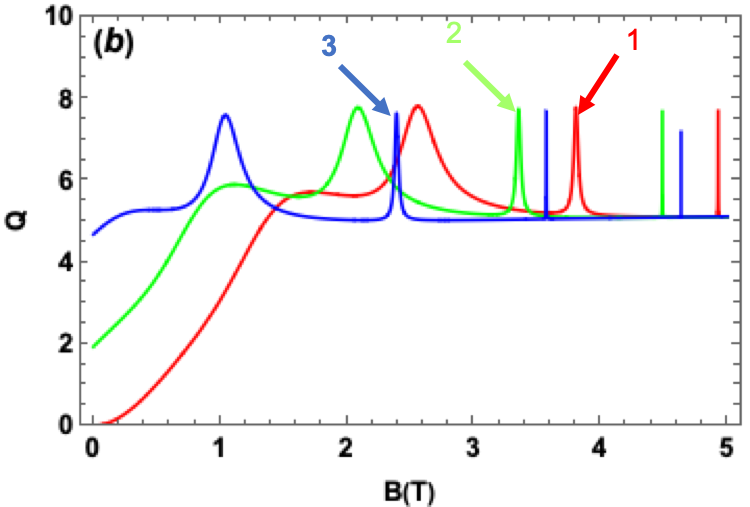}
\includegraphics[scale=0.46]{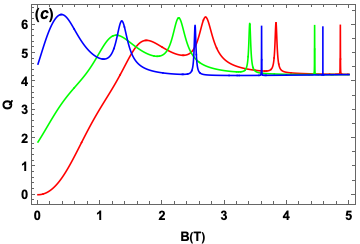}
		\caption{(color online) Scattering efficiency $Q$ as a function of magnetic field $B$ for AB-flux field $\mu = 0$ (red color), $\frac{1}{2}$ (green color), $\frac{3}{2}$ (blue color)  and three values of incident energy $E$. (a): $E=7$ meV, (b): $E=20$ meV and (c): $E=30$ meV.}
		\label{fig4art}
	\end{figure}

	In Fig. \ref{fig4art}, we now fix the radius at $R=50$ nm and the energy of the incident electron at values $E=7, 20, 30$ meV, and examine the scattering efficiency $Q$ as a function of the magnetic field $B$ with three values of magnetic flux $\mu= 0, 1/2, 3/2$, as depicted. For $E=7$ meV, Fig. \ref{fig4art}a  shows that at $B=0$ T, $Q$ is null in the absence of magnetic flux, but it starts to increase once  the flux is applied, and specifically, it takes the value $1$ at $\mu=3/2$. We also see that the higher the magnetic flux value, the larger the resonance peaks start to appear at smaller values of $B$. In Figs. \ref{fig4art}(b,c), where the energy of the incident electron $E$ is increased, the same conclusions are valid as seen in Fig. \ref{fig4art}a except that the minimum efficiency at $B=0$ T takes significant values until it reaches the value $4.6$ at $E=30$ meV and $\mu=3/2$, as displayed in Fig. \ref{fig4art}c.
	  
%

\begin{figure}[ht]
	\centering
	\includegraphics[scale=0.46]{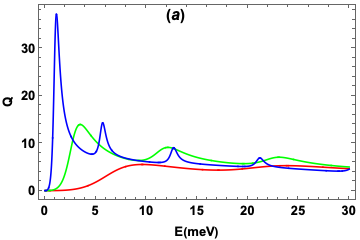}\includegraphics[scale=0.46]{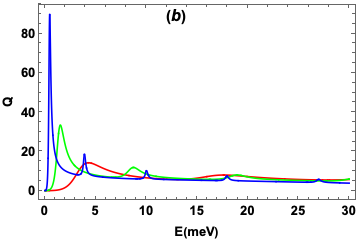}
	\includegraphics[scale=0.46]{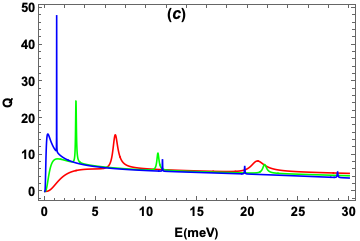}
	\caption{(color online) Scattering efficiency $Q$ as a function of incident energy $E$ for magnetic field $B=2.2$ T, radius $R=40$ (blue color), 50 (green color), 60 (red color) nm, and three values of magnetic flux (a): $\mu=0$, (b): $\mu=\frac{1}{2}$, (c): $\mu=\frac{3}{2}$.}
	\label{fig5art}
\end{figure}

Another analysis of the scattering phenomenon is shown in Fig. \ref{fig5art}, where we plot the scattering efficiency $Q$ as a function of the incident energy $E$ for a  $B=2.2$ T and three values of the GQD radius $R = 40, 50, 60$ nm, with magnetic flux (a): $\mu = $, (b): $\mu = 1/2$, (c): $\mu = 3/2$. In Fig. \ref{fig5art}, we see firstly that $Q$ is always zero at $E=0$ meV whatever the value of $\mu$. Secondly, we can clearly see that if we increase  $E$, $Q$ shows an oscillatory behavior with peaks of large amplitude at small values of  $E$. As long as  the energy $E$ is increased, these oscillations are damped until $Q$ takes a constant value, i.e., $Q\approx5$. A very important remark that can be drawn here is that the maximum value of  $Q$ increases with magnetic flux, as clearly indicated in Figs. \ref{fig5art}(a,b,c).

\begin{figure}[ht]
	\centering
\includegraphics[scale=0.45]{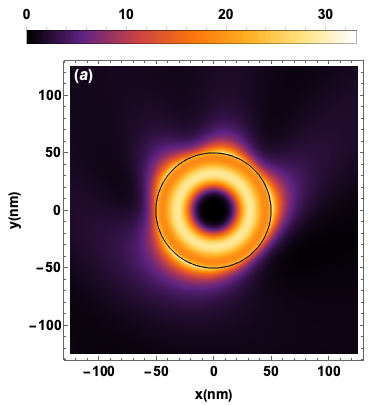}\includegraphics[scale=0.45]{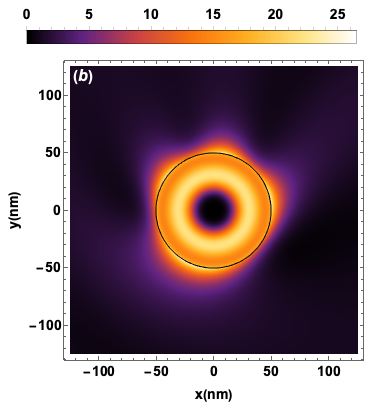}
\includegraphics[scale=0.45]{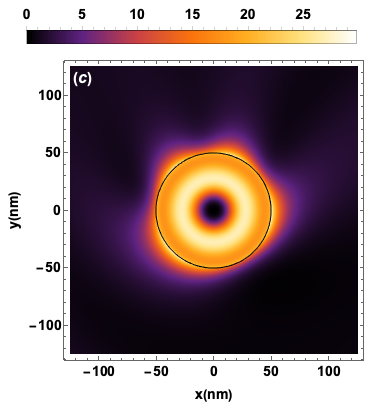}
	
	\caption{(color online) Density in real space corresponds to scattering mode $m=3$ for incident energy $E=20$ meV, radius $R=50$ nm, and different values of magnetic flux and magnetic field (a): ($\mu=0$, $B=3.8$ T), (b): ($\mu=0.5$, $B=3.34$ T),  (c): ($\mu=1.5$, $B=2.4$ T). The geographical GQD is indicated by the black circle.}
	\label{fig6art}
\end{figure}

\begin{figure}[ht]
	\centering
	\includegraphics[scale=0.46]{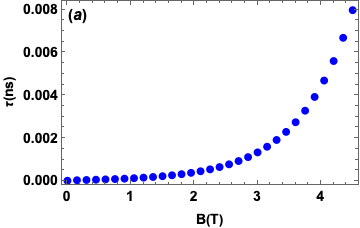}\includegraphics[scale=0.46]{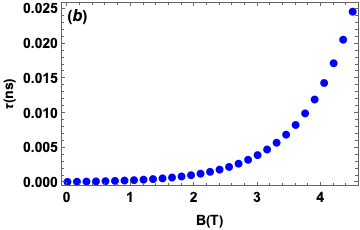}\includegraphics[scale=0.44]{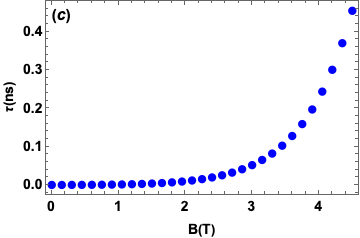}
	\includegraphics[scale=0.46]{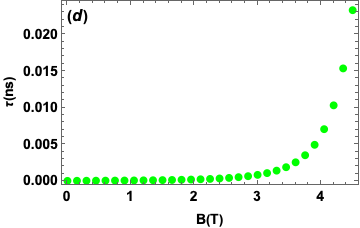}\includegraphics[scale=0.45]{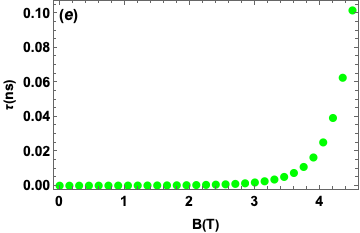}\includegraphics[scale=0.43]{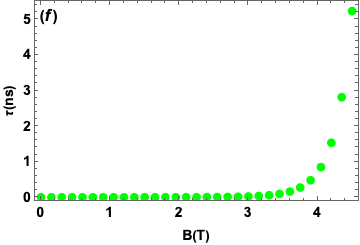}
	\caption{(color online) The trapping time $\tau$ as function of magnetic field  $B$ for two scattering modes $m=0$ (blue color), $m=3$ (green color), and  three values of magnetic flux (a,d): $\mu=0$, (b,e): $\mu=\frac{1}{2}$ and (c,f): $\mu=\frac{3}{2}$.}
	\label{fig7art}
\end{figure}

Next, we carry out another study of the scattering phenomenon based on the density in real space, for which we examine in Fig. \ref{fig6art} the density in the field near the GQD for an incident energy $E=20$ meV and three values of magnetic flux (a): $\mu=0$, (b): $\mu=1/$2, (c): $\mu=3/2$. We choose the scattering mode $m=3$ where scattering is made with a very sharp resonance peak, and consequently we expect notable electron trapping effects. The values of the magnetic field  $B$ are those corresponding to the peaks indicated by the labels $(1,2,3)$ in Fig. \ref{fig4art}b, the geometry of the GQD is indicated by the black circle. In Fig. \ref{fig6art}a with $B = 3.8$ T and zero magnetic flux, we see that most of the electron density is concentrated inside and at the boundary of the GQD with a high scattering efficiency. Figs. \ref{fig6art}(b,c) with ($B=3.34$ T, $\mu=1/2$), ($B=2.4$ T, $\mu=3/2$), respectively, show that the electron density inside the GQD is enhanced very clearly with the increase in magnetic flux and starts to form a very intense cloud around the center of the GQD, with an extremely high scattering efficiency, and consequently the probability of trapping the electron inside the GQD also becomes high.

   We focus on the two scattering modes $m=0$, where the scattering is non-resonant, and $m=3$, where the scattering has a very clear resonance peak (noticeable trapping effect), in order to study the effect of the magnetic flux on the trapping time of the electrons inside the GQD. The numerical solution of the transcendental equation \eqref{e:34} for each resonance allows us to determine the sets of values ($E_r$,$E_i$) as well as their corresponding magnetic field $B$ and consequently to deduce the trapping time $\tau$. Fig. \ref{fig7art} shows the trapping time $\tau$ as a function of the magnetic field  $B$ for a GQD radius $R=50$ nm, the two  modes ($m=0$ (blue color), $m=3$ (green color)), and the three  values of magnetic flux (first column: $\mu=0$, second: $\mu=1/2$, third: $\mu=3/2$). In general, we observe that the trapping time increases with the magnetic field $B$. In contrast to  $m=3$, we notice  that the trapping time in the $m=0$ mode begins to be visible at lower values of magnetic field $B$, and this is reinforced by what was found in \cite{pena2022electrons}. The most important observation of interest in our work is that the increase in the magnetic flux leads to a notable increase in the trapping time of the electrons inside the GQD. For example, at $B=4.5$ T, $\tau$ takes the following values: ($\tau=0.008$ ns at $\mu=0$, $\tau=0.0245$ ns at $\mu=1/2$ and $\tau=0.46$ ns at $\mu=3/2$) for $m=0$ and ($\tau=0.023$ ns at $\mu=0$, $\tau=0.10$ ns at $\mu=1/2$ and $\tau=6$ ns at $\mu=3/2$) for $m=3$. 

	\section{Conclusion} \label{cc}
	
		In summary, we carried out a theoretical study of the elastic diffusion of electrons in magnetic graphene quantum dots (GQDs) subjected to a magnetic flux. We showed that the influence of the magnetic flux on improving the scattering efficiency and, in particular, the trapping time of quasi-bound states that can be induced in GQDs immersed in a homogeneous external magnetic field. To this end, we have developed a theoretical model that describes the behavior of Dirac fermions that allowed us to achieve our main objective. To treat the problem in detail, we first solved the Dirac equation analytically to determine the eigenspinors. Then, we carried out scattering analysis. In this respect, we defined the physical quantities used in our study to investigate scattering, such as the scattering coefficients, the probability density, the scattering efficiency, and the trapping time.

	On the basis of the numerical results, we investigated the GQD system under several values of the physical parameters: the incident electron energy, the magnetic field intensity, the GQD radius, the angular momentum, and the magnetic flux. Indeed, we found that with increased magnetic flux, the scattering efficiency takes significant non-zero values at zero magnetic field, and the quasi-bound states also start to become measurable at smaller values of the GQD radius. The magnetic flux can be considered an important parameter to control scattering and excitation of the quasi-bound states. Secondly, in terms of the density, we have shown that it increases inside the GQD with an increasing magnetic flux. As a result, the probability of trapping the electrons inside the GQD becomes very high. Finally, in terms of complex incident energy, we showed that a significant increase in the electron trapping time inside the GQDs can be acheived with an increase in the magnetic flux.

	\bibliography{Ref.bib}

\begin{thebibliography}{33}%
\makeatletter
\providecommand \@ifxundefined [1]{%
 \@ifx{#1\undefined}
}%
\providecommand \@ifnum [1]{%
 \ifnum #1\expandafter \@firstoftwo
 \else \expandafter \@secondoftwo
 \fi
}%
\providecommand \@ifx [1]{%
 \ifx #1\expandafter \@firstoftwo
 \else \expandafter \@secondoftwo
 \fi
}%
\providecommand \natexlab [1]{#1}%
\providecommand \enquote  [1]{``#1''}%
\providecommand \bibnamefont  [1]{#1}%
\providecommand \bibfnamefont [1]{#1}%
\providecommand \citenamefont [1]{#1}%
\providecommand \href@noop [0]{\@secondoftwo}%
\providecommand \href [0]{\begingroup \@sanitize@url \@href}%
\providecommand \@href[1]{\@@startlink{#1}\@@href}%
\providecommand \@@href[1]{\endgroup#1\@@endlink}%
\providecommand \@sanitize@url [0]{\catcode `\\12\catcode `\$12\catcode
  `\&12\catcode `\#12\catcode `\^12\catcode `\_12\catcode `\%12\relax}%
\providecommand \@@startlink[1]{}%
\providecommand \@@endlink[0]{}%
\providecommand \url  [0]{\begingroup\@sanitize@url \@url }%
\providecommand \@url [1]{\endgroup\@href {#1}{\urlprefix }}%
\providecommand \urlprefix  [0]{URL }%
\providecommand \Eprint [0]{\href }%
\providecommand \doibase [0]{https://doi.org/}%
\providecommand \selectlanguage [0]{\@gobble}%
\providecommand \bibinfo  [0]{\@secondoftwo}%
\providecommand \bibfield  [0]{\@secondoftwo}%
\providecommand \translation [1]{[#1]}%
\providecommand \BibitemOpen [0]{}%
\providecommand \bibitemStop [0]{}%
\providecommand \bibitemNoStop [0]{.\EOS\space}%
\providecommand \EOS [0]{\spacefactor3000\relax}%
\providecommand \BibitemShut  [1]{\csname bibitem#1\endcsname}%
\let\auto@bib@innerbib\@empty
\bibitem [{\citenamefont {Novoselov}\ \emph {et~al.}(2005)\citenamefont
  {Novoselov}, \citenamefont {Geim}, \citenamefont {Morozov}, \citenamefont
  {Jiang}, \citenamefont {Katsnelson}, \citenamefont {Grigorieva},
  \citenamefont {Dubonos}, \citenamefont {Firsov},\ and\ \citenamefont
  {AA}}]{novoselov2005two}%
  \BibitemOpen
  \bibfield  {author} {\bibinfo {author} {\bibfnamefont {K.~S.}\ \bibnamefont
  {Novoselov}}, \bibinfo {author} {\bibfnamefont {A.~K.}\ \bibnamefont {Geim}},
  \bibinfo {author} {\bibfnamefont {S.~V.}\ \bibnamefont {Morozov}}, \bibinfo
  {author} {\bibfnamefont {D.}~\bibnamefont {Jiang}}, \bibinfo {author}
  {\bibfnamefont {M.~I.}\ \bibnamefont {Katsnelson}}, \bibinfo {author}
  {\bibfnamefont {I.~V.}\ \bibnamefont {Grigorieva}}, \bibinfo {author}
  {\bibfnamefont {S.}~\bibnamefont {Dubonos}}, \bibinfo {author} {\bibnamefont
  {Firsov}},\ and\ \bibinfo {author} {\bibnamefont {AA}},\ }\bibfield  {title}
  {\bibinfo {title} {Two-dimensional gas of massless dirac fermions in
  graphene},\ }\href@noop {} {\bibfield  {journal} {\bibinfo  {journal}
  {nature}\ }\textbf {\bibinfo {volume} {438}},\ \bibinfo {pages} {197}
  (\bibinfo {year} {2005})}\BibitemShut {NoStop}%
\bibitem [{\citenamefont {Neto}\ \emph {et~al.}(2009)\citenamefont {Neto},
  \citenamefont {Guinea}, \citenamefont {Peres}, \citenamefont {Novoselov},\
  and\ \citenamefont {Geim}}]{neto2009electronic}%
  \BibitemOpen
  \bibfield  {author} {\bibinfo {author} {\bibfnamefont {A.~C.}\ \bibnamefont
  {Neto}}, \bibinfo {author} {\bibfnamefont {F.}~\bibnamefont {Guinea}},
  \bibinfo {author} {\bibfnamefont {N.~M.}\ \bibnamefont {Peres}}, \bibinfo
  {author} {\bibfnamefont {K.~S.}\ \bibnamefont {Novoselov}},\ and\ \bibinfo
  {author} {\bibfnamefont {A.~K.}\ \bibnamefont {Geim}},\ }\bibfield  {title}
  {\bibinfo {title} {The electronic properties of graphene},\ }\href@noop {}
  {\bibfield  {journal} {\bibinfo  {journal} {Reviews of modern physics}\
  }\textbf {\bibinfo {volume} {81}},\ \bibinfo {pages} {109} (\bibinfo {year}
  {2009})}\BibitemShut {NoStop}%
\bibitem [{\citenamefont {Goerbig}\ \emph {et~al.}(2006)\citenamefont
  {Goerbig}, \citenamefont {Moessner},\ and\ \citenamefont
  {Dou{\c{c}}ot}}]{goerbig2006electron}%
  \BibitemOpen
  \bibfield  {author} {\bibinfo {author} {\bibfnamefont {M.~O.}\ \bibnamefont
  {Goerbig}}, \bibinfo {author} {\bibfnamefont {R.}~\bibnamefont {Moessner}},\
  and\ \bibinfo {author} {\bibfnamefont {B.}~\bibnamefont {Dou{\c{c}}ot}},\
  }\bibfield  {title} {\bibinfo {title} {Electron interactions in graphene in a
  strong magnetic field},\ }\href@noop {} {\bibfield  {journal} {\bibinfo
  {journal} {Physical Review B}\ }\textbf {\bibinfo {volume} {74}},\ \bibinfo
  {pages} {161407} (\bibinfo {year} {2006})}\BibitemShut {NoStop}%
\bibitem [{\citenamefont {Avetisyan}\ \emph {et~al.}(2009)\citenamefont
  {Avetisyan}, \citenamefont {Partoens},\ and\ \citenamefont
  {Peeters}}]{avetisyan2009electric}%
  \BibitemOpen
  \bibfield  {author} {\bibinfo {author} {\bibfnamefont {A.}~\bibnamefont
  {Avetisyan}}, \bibinfo {author} {\bibfnamefont {B.}~\bibnamefont
  {Partoens}},\ and\ \bibinfo {author} {\bibfnamefont {F.}~\bibnamefont
  {Peeters}},\ }\bibfield  {title} {\bibinfo {title} {Electric field tuning of
  the band gap in graphene multilayers},\ }\href@noop {} {\bibfield  {journal}
  {\bibinfo  {journal} {Physical Review B}\ }\textbf {\bibinfo {volume} {79}},\
  \bibinfo {pages} {035421} (\bibinfo {year} {2009})}\BibitemShut {NoStop}%
\bibitem [{\citenamefont {Zhang}\ \emph {et~al.}(2005)\citenamefont {Zhang},
  \citenamefont {Tan}, \citenamefont {Stormer},\ and\ \citenamefont
  {Kim}}]{zhang2005experimental}%
  \BibitemOpen
  \bibfield  {author} {\bibinfo {author} {\bibfnamefont {Y.}~\bibnamefont
  {Zhang}}, \bibinfo {author} {\bibfnamefont {Y.-W.}\ \bibnamefont {Tan}},
  \bibinfo {author} {\bibfnamefont {H.~L.}\ \bibnamefont {Stormer}},\ and\
  \bibinfo {author} {\bibfnamefont {P.}~\bibnamefont {Kim}},\ }\bibfield
  {title} {\bibinfo {title} {Experimental observation of the quantum hall
  effect and berry's phase in graphene},\ }\href@noop {} {\bibfield  {journal}
  {\bibinfo  {journal} {nature}\ }\textbf {\bibinfo {volume} {438}},\ \bibinfo
  {pages} {201} (\bibinfo {year} {2005})}\BibitemShut {NoStop}%
\bibitem [{\citenamefont {Jiang}\ \emph {et~al.}(2007)\citenamefont {Jiang},
  \citenamefont {Zhang}, \citenamefont {Tan}, \citenamefont {Stormer},\ and\
  \citenamefont {Kim}}]{jiang2007quantum}%
  \BibitemOpen
  \bibfield  {author} {\bibinfo {author} {\bibfnamefont {Z.}~\bibnamefont
  {Jiang}}, \bibinfo {author} {\bibfnamefont {Y.}~\bibnamefont {Zhang}},
  \bibinfo {author} {\bibfnamefont {Y.-W.}\ \bibnamefont {Tan}}, \bibinfo
  {author} {\bibfnamefont {H.}~\bibnamefont {Stormer}},\ and\ \bibinfo {author}
  {\bibfnamefont {P.}~\bibnamefont {Kim}},\ }\bibfield  {title} {\bibinfo
  {title} {Quantum hall effect in graphene},\ }\href@noop {} {\bibfield
  {journal} {\bibinfo  {journal} {Solid state communications}\ }\textbf
  {\bibinfo {volume} {143}},\ \bibinfo {pages} {14} (\bibinfo {year}
  {2007})}\BibitemShut {NoStop}%
\bibitem [{\citenamefont {Barlas}\ \emph {et~al.}(2012)\citenamefont {Barlas},
  \citenamefont {Yang},\ and\ \citenamefont {MacDonald}}]{barlas2012quantum}%
  \BibitemOpen
  \bibfield  {author} {\bibinfo {author} {\bibfnamefont {Y.}~\bibnamefont
  {Barlas}}, \bibinfo {author} {\bibfnamefont {K.}~\bibnamefont {Yang}},\ and\
  \bibinfo {author} {\bibfnamefont {A.}~\bibnamefont {MacDonald}},\ }\bibfield
  {title} {\bibinfo {title} {Quantum hall effects in graphene-based
  two-dimensional electron systems},\ }\href@noop {} {\bibfield  {journal}
  {\bibinfo  {journal} {Nanotechnology}\ }\textbf {\bibinfo {volume} {23}},\
  \bibinfo {pages} {052001} (\bibinfo {year} {2012})}\BibitemShut {NoStop}%
\bibitem [{\citenamefont {Recher}\ \emph {et~al.}(2007)\citenamefont {Recher},
  \citenamefont {Trauzettel}, \citenamefont {Rycerz}, \citenamefont {Blanter},
  \citenamefont {Beenakker},\ and\ \citenamefont
  {Morpurgo}}]{recher2007aharonov}%
  \BibitemOpen
  \bibfield  {author} {\bibinfo {author} {\bibfnamefont {P.}~\bibnamefont
  {Recher}}, \bibinfo {author} {\bibfnamefont {B.}~\bibnamefont {Trauzettel}},
  \bibinfo {author} {\bibfnamefont {A.}~\bibnamefont {Rycerz}}, \bibinfo
  {author} {\bibfnamefont {Y.~M.}\ \bibnamefont {Blanter}}, \bibinfo {author}
  {\bibfnamefont {C.}~\bibnamefont {Beenakker}},\ and\ \bibinfo {author}
  {\bibfnamefont {A.}~\bibnamefont {Morpurgo}},\ }\bibfield  {title} {\bibinfo
  {title} {Aharonov-bohm effect and broken valley degeneracy in graphene
  rings},\ }\href@noop {} {\bibfield  {journal} {\bibinfo  {journal} {Physical
  Review B}\ }\textbf {\bibinfo {volume} {76}},\ \bibinfo {pages} {235404}
  (\bibinfo {year} {2007})}\BibitemShut {NoStop}%
\bibitem [{\citenamefont {Jackiw}\ \emph {et~al.}(2009)\citenamefont {Jackiw},
  \citenamefont {Milstein}, \citenamefont {Pi},\ and\ \citenamefont
  {Terekhov}}]{jackiw2009induced}%
  \BibitemOpen
  \bibfield  {author} {\bibinfo {author} {\bibfnamefont {R.}~\bibnamefont
  {Jackiw}}, \bibinfo {author} {\bibfnamefont {A.}~\bibnamefont {Milstein}},
  \bibinfo {author} {\bibfnamefont {S.-Y.}\ \bibnamefont {Pi}},\ and\ \bibinfo
  {author} {\bibfnamefont {I.}~\bibnamefont {Terekhov}},\ }\bibfield  {title}
  {\bibinfo {title} {Induced current and aharonov-bohm effect in graphene},\
  }\href@noop {} {\bibfield  {journal} {\bibinfo  {journal} {Physical Review
  B}\ }\textbf {\bibinfo {volume} {80}},\ \bibinfo {pages} {033413} (\bibinfo
  {year} {2009})}\BibitemShut {NoStop}%
\bibitem [{\citenamefont {Schelter}\ \emph {et~al.}(2012)\citenamefont
  {Schelter}, \citenamefont {Recher},\ and\ \citenamefont
  {Trauzettel}}]{schelter2012aharonov}%
  \BibitemOpen
  \bibfield  {author} {\bibinfo {author} {\bibfnamefont {J.}~\bibnamefont
  {Schelter}}, \bibinfo {author} {\bibfnamefont {P.}~\bibnamefont {Recher}},\
  and\ \bibinfo {author} {\bibfnamefont {B.}~\bibnamefont {Trauzettel}},\
  }\bibfield  {title} {\bibinfo {title} {The aharonov--bohm effect in graphene
  rings},\ }\href@noop {} {\bibfield  {journal} {\bibinfo  {journal} {Solid
  state communications}\ }\textbf {\bibinfo {volume} {152}},\ \bibinfo {pages}
  {1411} (\bibinfo {year} {2012})}\BibitemShut {NoStop}%
\bibitem [{\citenamefont {Luican}\ \emph {et~al.}(2011)\citenamefont {Luican},
  \citenamefont {Li},\ and\ \citenamefont {Andrei}}]{luican2011quantized}%
  \BibitemOpen
  \bibfield  {author} {\bibinfo {author} {\bibfnamefont {A.}~\bibnamefont
  {Luican}}, \bibinfo {author} {\bibfnamefont {G.}~\bibnamefont {Li}},\ and\
  \bibinfo {author} {\bibfnamefont {E.~Y.}\ \bibnamefont {Andrei}},\ }\bibfield
   {title} {\bibinfo {title} {Quantized landau level spectrum and its density
  dependence in graphene},\ }\href@noop {} {\bibfield  {journal} {\bibinfo
  {journal} {Physical Review B}\ }\textbf {\bibinfo {volume} {83}},\ \bibinfo
  {pages} {041405} (\bibinfo {year} {2011})}\BibitemShut {NoStop}%
\bibitem [{\citenamefont {Yin}\ \emph {et~al.}(2017)\citenamefont {Yin},
  \citenamefont {Bai}, \citenamefont {Wang}, \citenamefont {Li}, \citenamefont
  {Zhang},\ and\ \citenamefont {He}}]{yin2017landau}%
  \BibitemOpen
  \bibfield  {author} {\bibinfo {author} {\bibfnamefont {L.-J.}\ \bibnamefont
  {Yin}}, \bibinfo {author} {\bibfnamefont {K.-K.}\ \bibnamefont {Bai}},
  \bibinfo {author} {\bibfnamefont {W.-X.}\ \bibnamefont {Wang}}, \bibinfo
  {author} {\bibfnamefont {S.-Y.}\ \bibnamefont {Li}}, \bibinfo {author}
  {\bibfnamefont {Y.}~\bibnamefont {Zhang}},\ and\ \bibinfo {author}
  {\bibfnamefont {L.}~\bibnamefont {He}},\ }\bibfield  {title} {\bibinfo
  {title} {Landau quantization of dirac fermions in graphene and its
  multilayers},\ }\href@noop {} {\bibfield  {journal} {\bibinfo  {journal}
  {Frontiers of Physics}\ }\textbf {\bibinfo {volume} {12}},\ \bibinfo {pages}
  {1} (\bibinfo {year} {2017})}\BibitemShut {NoStop}%
\bibitem [{\citenamefont {Nemec}\ and\ \citenamefont
  {Cuniberti}(2007)}]{nemec2007hofstadter}%
  \BibitemOpen
  \bibfield  {author} {\bibinfo {author} {\bibfnamefont {N.}~\bibnamefont
  {Nemec}}\ and\ \bibinfo {author} {\bibfnamefont {G.}~\bibnamefont
  {Cuniberti}},\ }\bibfield  {title} {\bibinfo {title} {Hofstadter butterflies
  of bilayer graphene},\ }\href@noop {} {\bibfield  {journal} {\bibinfo
  {journal} {Physical Review B}\ }\textbf {\bibinfo {volume} {75}},\ \bibinfo
  {pages} {201404} (\bibinfo {year} {2007})}\BibitemShut {NoStop}%
\bibitem [{\citenamefont {Beenakker}(2008)}]{beenakker2008colloquium}%
  \BibitemOpen
  \bibfield  {author} {\bibinfo {author} {\bibfnamefont {C.}~\bibnamefont
  {Beenakker}},\ }\bibfield  {title} {\bibinfo {title} {Colloquium: Andreev
  reflection and klein tunneling in graphene},\ }\href@noop {} {\bibfield
  {journal} {\bibinfo  {journal} {Reviews of Modern Physics}\ }\textbf
  {\bibinfo {volume} {80}},\ \bibinfo {pages} {1337} (\bibinfo {year}
  {2008})}\BibitemShut {NoStop}%
\bibitem [{\citenamefont {Allain}\ and\ \citenamefont
  {Fuchs}(2011)}]{allain2011klein}%
  \BibitemOpen
  \bibfield  {author} {\bibinfo {author} {\bibfnamefont {P.~E.}\ \bibnamefont
  {Allain}}\ and\ \bibinfo {author} {\bibfnamefont {J.-N.}\ \bibnamefont
  {Fuchs}},\ }\bibfield  {title} {\bibinfo {title} {Klein tunneling in
  graphene: optics with massless electrons},\ }\href@noop {} {\bibfield
  {journal} {\bibinfo  {journal} {The European Physical Journal B}\ }\textbf
  {\bibinfo {volume} {83}},\ \bibinfo {pages} {301} (\bibinfo {year}
  {2011})}\BibitemShut {NoStop}%
\bibitem [{\citenamefont {Peres}\ \emph {et~al.}(2006)\citenamefont {Peres},
  \citenamefont {Neto},\ and\ \citenamefont {Guinea}}]{peres2006dirac}%
  \BibitemOpen
  \bibfield  {author} {\bibinfo {author} {\bibfnamefont {N.}~\bibnamefont
  {Peres}}, \bibinfo {author} {\bibfnamefont {A.~C.}\ \bibnamefont {Neto}},\
  and\ \bibinfo {author} {\bibfnamefont {F.}~\bibnamefont {Guinea}},\
  }\bibfield  {title} {\bibinfo {title} {Dirac fermion confinement in
  graphene},\ }\href@noop {} {\bibfield  {journal} {\bibinfo  {journal}
  {Physical Review B}\ }\textbf {\bibinfo {volume} {73}},\ \bibinfo {pages}
  {241403} (\bibinfo {year} {2006})}\BibitemShut {NoStop}%
\bibitem [{\citenamefont {Chen}\ \emph {et~al.}(2007)\citenamefont {Chen},
  \citenamefont {Apalkov},\ and\ \citenamefont {Chakraborty}}]{chen2007fock}%
  \BibitemOpen
  \bibfield  {author} {\bibinfo {author} {\bibfnamefont {H.-Y.}\ \bibnamefont
  {Chen}}, \bibinfo {author} {\bibfnamefont {V.}~\bibnamefont {Apalkov}},\ and\
  \bibinfo {author} {\bibfnamefont {T.}~\bibnamefont {Chakraborty}},\
  }\bibfield  {title} {\bibinfo {title} {Fock-darwin states of dirac electrons
  in graphene-based artificial atoms},\ }\href@noop {} {\bibfield  {journal}
  {\bibinfo  {journal} {Physical review letters}\ }\textbf {\bibinfo {volume}
  {98}},\ \bibinfo {pages} {186803} (\bibinfo {year} {2007})}\BibitemShut
  {NoStop}%
\bibitem [{\citenamefont {Silvestrov}\ and\ \citenamefont
  {Efetov}(2007)}]{silvestrov2007quantum}%
  \BibitemOpen
  \bibfield  {author} {\bibinfo {author} {\bibfnamefont {P.}~\bibnamefont
  {Silvestrov}}\ and\ \bibinfo {author} {\bibfnamefont {K.}~\bibnamefont
  {Efetov}},\ }\bibfield  {title} {\bibinfo {title} {Quantum dots in
  graphene},\ }\href@noop {} {\bibfield  {journal} {\bibinfo  {journal}
  {Physical review letters}\ }\textbf {\bibinfo {volume} {98}},\ \bibinfo
  {pages} {016802} (\bibinfo {year} {2007})}\BibitemShut {NoStop}%
\bibitem [{\citenamefont {Fehske}\ \emph {et~al.}(2015)\citenamefont {Fehske},
  \citenamefont {Hager},\ and\ \citenamefont {Pieper}}]{fehske2015electron}%
  \BibitemOpen
  \bibfield  {author} {\bibinfo {author} {\bibfnamefont {H.}~\bibnamefont
  {Fehske}}, \bibinfo {author} {\bibfnamefont {G.}~\bibnamefont {Hager}},\ and\
  \bibinfo {author} {\bibfnamefont {A.}~\bibnamefont {Pieper}},\ }\bibfield
  {title} {\bibinfo {title} {Electron confinement in graphene with gate-defined
  quantum dots},\ }\href@noop {} {\bibfield  {journal} {\bibinfo  {journal}
  {physica status solidi (b)}\ }\textbf {\bibinfo {volume} {252}},\ \bibinfo
  {pages} {1868} (\bibinfo {year} {2015})}\BibitemShut {NoStop}%
\bibitem [{\citenamefont {El~Azar}\ \emph {et~al.}()\citenamefont {El~Azar},
  \citenamefont {Jellal},\ and\ \citenamefont
  {Bouhlal}}]{jellal4358814electrons}%
  \BibitemOpen
  \bibfield  {author} {\bibinfo {author} {\bibfnamefont {M.}~\bibnamefont
  {El~Azar}}, \bibinfo {author} {\bibfnamefont {A.}~\bibnamefont {Jellal}},\
  and\ \bibinfo {author} {\bibfnamefont {A.}~\bibnamefont {Bouhlal}},\
  }\bibfield  {title} {\bibinfo {title} {Electrons trapped in magnetic
  controlled gapped graphene-based quantum dots},\ }\href@noop {} {\bibinfo
  {journal} {Available at SSRN 4358814}\ }\BibitemShut {NoStop}%
\bibitem [{\citenamefont {Pena}(2022{\natexlab{a}})}]{pena2022electron}%
  \BibitemOpen
\bibfield  {journal} {  }\bibfield  {author} {\bibinfo {author} {\bibfnamefont
  {A.}~\bibnamefont {Pena}},\ }\bibfield  {title} {\bibinfo {title} {Electron
  trapping in twisted light driven graphene quantum dots},\ }\href@noop {}
  {\bibfield  {journal} {\bibinfo  {journal} {Physical Review B}\ }\textbf
  {\bibinfo {volume} {105}},\ \bibinfo {pages} {045405} (\bibinfo {year}
  {2022}{\natexlab{a}})}\BibitemShut {NoStop}%
\bibitem [{\citenamefont {De~Martino}\ \emph {et~al.}(2007)\citenamefont
  {De~Martino}, \citenamefont {Dell’Anna},\ and\ \citenamefont
  {Egger}}]{de2007magnetic}%
  \BibitemOpen
  \bibfield  {author} {\bibinfo {author} {\bibfnamefont {A.}~\bibnamefont
  {De~Martino}}, \bibinfo {author} {\bibfnamefont {L.}~\bibnamefont
  {Dell’Anna}},\ and\ \bibinfo {author} {\bibfnamefont {R.}~\bibnamefont
  {Egger}},\ }\bibfield  {title} {\bibinfo {title} {Magnetic confinement of
  massless dirac fermions in graphene},\ }\href@noop {} {\bibfield  {journal}
  {\bibinfo  {journal} {Physical review letters}\ }\textbf {\bibinfo {volume}
  {98}},\ \bibinfo {pages} {066802} (\bibinfo {year} {2007})}\BibitemShut
  {NoStop}%
\bibitem [{\citenamefont {Wang}\ and\ \citenamefont
  {Jin}(2009)}]{wang2009bound}%
  \BibitemOpen
  \bibfield  {author} {\bibinfo {author} {\bibfnamefont {D.}~\bibnamefont
  {Wang}}\ and\ \bibinfo {author} {\bibfnamefont {G.}~\bibnamefont {Jin}},\
  }\bibfield  {title} {\bibinfo {title} {Bound states of dirac electrons in a
  graphene-based magnetic quantum dot},\ }\href@noop {} {\bibfield  {journal}
  {\bibinfo  {journal} {Physics Letters A}\ }\textbf {\bibinfo {volume}
  {373}},\ \bibinfo {pages} {4082} (\bibinfo {year} {2009})}\BibitemShut
  {NoStop}%
\bibitem [{\citenamefont {Pan}\ \emph {et~al.}(2020)\citenamefont {Pan},
  \citenamefont {Ji}, \citenamefont {Li},\ and\ \citenamefont
  {Liu}}]{pan2020quasi}%
  \BibitemOpen
  \bibfield  {author} {\bibinfo {author} {\bibfnamefont {Y.}~\bibnamefont
  {Pan}}, \bibinfo {author} {\bibfnamefont {H.}~\bibnamefont {Ji}}, \bibinfo
  {author} {\bibfnamefont {X.-Q.}\ \bibnamefont {Li}},\ and\ \bibinfo {author}
  {\bibfnamefont {H.}~\bibnamefont {Liu}},\ }\bibfield  {title} {\bibinfo
  {title} {Quasi-bound states in an npn-type nanometer-scale graphene quantum
  dot under a magnetic field},\ }\href@noop {} {\bibfield  {journal} {\bibinfo
  {journal} {Scientific Reports}\ }\textbf {\bibinfo {volume} {10}},\ \bibinfo
  {pages} {20426} (\bibinfo {year} {2020})}\BibitemShut {NoStop}%
\bibitem [{\citenamefont {Pena}(2022{\natexlab{b}})}]{pena2022lifetime}%
  \BibitemOpen
  \bibfield  {author} {\bibinfo {author} {\bibfnamefont {A.}~\bibnamefont
  {Pena}},\ }\bibfield  {title} {\bibinfo {title} {Lifetime enhancement of
  quasibound states in graphene quantum dots via circularly polarized light},\
  }\href@noop {} {\bibfield  {journal} {\bibinfo  {journal} {Physical Review
  B}\ }\textbf {\bibinfo {volume} {105}},\ \bibinfo {pages} {125408} (\bibinfo
  {year} {2022}{\natexlab{b}})}\BibitemShut {NoStop}%
\bibitem [{\citenamefont {Pena}(2022{\natexlab{c}})}]{pena2022electrons}%
  \BibitemOpen
  \bibfield  {author} {\bibinfo {author} {\bibfnamefont {A.}~\bibnamefont
  {Pena}},\ }\bibfield  {title} {\bibinfo {title} {Electron trapping in
  magnetic driven graphene quantum dots},\ }\href@noop {} {\bibfield  {journal}
  {\bibinfo  {journal} {Physica E: Low-dimensional Systems and Nanostructures}\
  }\textbf {\bibinfo {volume} {141}},\ \bibinfo {pages} {115245} (\bibinfo
  {year} {2022}{\natexlab{c}})}\BibitemShut {NoStop}%
\bibitem [{\citenamefont {Ikhdair}\ \emph {et~al.}(2015)\citenamefont
  {Ikhdair}, \citenamefont {Falaye},\ and\ \citenamefont
  {Hamzavi}}]{ikhdair2015nonrelativistic}%
  \BibitemOpen
  \bibfield  {author} {\bibinfo {author} {\bibfnamefont {S.~M.}\ \bibnamefont
  {Ikhdair}}, \bibinfo {author} {\bibfnamefont {B.~J.}\ \bibnamefont
  {Falaye}},\ and\ \bibinfo {author} {\bibfnamefont {M.}~\bibnamefont
  {Hamzavi}},\ }\bibfield  {title} {\bibinfo {title} {Nonrelativistic molecular
  models under external magnetic and ab flux fields},\ }\href@noop {}
  {\bibfield  {journal} {\bibinfo  {journal} {Annals of Physics}\ }\textbf
  {\bibinfo {volume} {353}},\ \bibinfo {pages} {282} (\bibinfo {year}
  {2015})}\BibitemShut {NoStop}%
\bibitem [{\citenamefont {Hewageegana}\ and\ \citenamefont
  {Apalkov}(2008)}]{hewageegana2008electron}%
  \BibitemOpen
  \bibfield  {author} {\bibinfo {author} {\bibfnamefont {P.}~\bibnamefont
  {Hewageegana}}\ and\ \bibinfo {author} {\bibfnamefont {V.}~\bibnamefont
  {Apalkov}},\ }\bibfield  {title} {\bibinfo {title} {Electron localization in
  graphene quantum dots},\ }\href@noop {} {\bibfield  {journal} {\bibinfo
  {journal} {Physical Review B}\ }\textbf {\bibinfo {volume} {77}},\ \bibinfo
  {pages} {245426} (\bibinfo {year} {2008})}\BibitemShut {NoStop}%
\bibitem [{\citenamefont {Cserti}\ \emph {et~al.}(2007)\citenamefont {Cserti},
  \citenamefont {P{\'a}lyi},\ and\ \citenamefont
  {P{\'e}terfalvi}}]{cserti2007caustics}%
  \BibitemOpen
  \bibfield  {author} {\bibinfo {author} {\bibfnamefont {J.}~\bibnamefont
  {Cserti}}, \bibinfo {author} {\bibfnamefont {A.}~\bibnamefont {P{\'a}lyi}},\
  and\ \bibinfo {author} {\bibfnamefont {C.}~\bibnamefont {P{\'e}terfalvi}},\
  }\bibfield  {title} {\bibinfo {title} {Caustics due to a negative refractive
  index in circular graphene p- n junctions},\ }\href@noop {} {\bibfield
  {journal} {\bibinfo  {journal} {Physical review letters}\ }\textbf {\bibinfo
  {volume} {99}},\ \bibinfo {pages} {246801} (\bibinfo {year}
  {2007})}\BibitemShut {NoStop}%
\bibitem [{\citenamefont {Heinisch}\ \emph {et~al.}(2013)\citenamefont
  {Heinisch}, \citenamefont {Bronold},\ and\ \citenamefont
  {Fehske}}]{heinisch2013mie}%
  \BibitemOpen
  \bibfield  {author} {\bibinfo {author} {\bibfnamefont {R.}~\bibnamefont
  {Heinisch}}, \bibinfo {author} {\bibfnamefont {F.}~\bibnamefont {Bronold}},\
  and\ \bibinfo {author} {\bibfnamefont {H.}~\bibnamefont {Fehske}},\
  }\bibfield  {title} {\bibinfo {title} {Mie scattering analog in graphene:
  Lensing, particle confinement, and depletion of klein tunneling},\
  }\href@noop {} {\bibfield  {journal} {\bibinfo  {journal} {Physical Review
  B}\ }\textbf {\bibinfo {volume} {87}},\ \bibinfo {pages} {155409} (\bibinfo
  {year} {2013})}\BibitemShut {NoStop}%
\bibitem [{\citenamefont {Schulz}\ \emph {et~al.}(2014)\citenamefont {Schulz},
  \citenamefont {Heinisch},\ and\ \citenamefont {Fehske}}]{schulz2014electron}%
  \BibitemOpen
  \bibfield  {author} {\bibinfo {author} {\bibfnamefont {C.}~\bibnamefont
  {Schulz}}, \bibinfo {author} {\bibfnamefont {R.}~\bibnamefont {Heinisch}},\
  and\ \bibinfo {author} {\bibfnamefont {H.}~\bibnamefont {Fehske}},\
  }\bibfield  {title} {\bibinfo {title} {Electron flow in circular graphene
  quantum dots},\ }\href@noop {} {\bibfield  {journal} {\bibinfo  {journal}
  {arXiv preprint arXiv:1412.2539}\ } (\bibinfo {year} {2014})}\BibitemShut
  {NoStop}%
\bibitem [{\citenamefont {Belokda}\ \emph {et~al.}(2022)\citenamefont
  {Belokda}, \citenamefont {Jellal},\ and\ \citenamefont
  {Atmani}}]{belokda2022electron}%
  \BibitemOpen
  \bibfield  {author} {\bibinfo {author} {\bibfnamefont {F.}~\bibnamefont
  {Belokda}}, \bibinfo {author} {\bibfnamefont {A.}~\bibnamefont {Jellal}},\
  and\ \bibinfo {author} {\bibfnamefont {E.~H.}\ \bibnamefont {Atmani}},\
  }\bibfield  {title} {\bibinfo {title} {Electron scattering of inhomogeneous
  gap in graphene quantum dots},\ }\href@noop {} {\bibfield  {journal}
  {\bibinfo  {journal} {Physics Letters A}\ }\textbf {\bibinfo {volume}
  {448}},\ \bibinfo {pages} {128325} (\bibinfo {year} {2022})}\BibitemShut
  {NoStop}%
\bibitem [{\citenamefont {Narimanov}\ \emph {et~al.}(1999)\citenamefont
  {Narimanov}, \citenamefont {Hackenbroich}, \citenamefont {Jacquod},\ and\
  \citenamefont {Stone}}]{narimanov1999semiclassical}%
  \BibitemOpen
  \bibfield  {author} {\bibinfo {author} {\bibfnamefont {E.~E.}\ \bibnamefont
  {Narimanov}}, \bibinfo {author} {\bibfnamefont {G.}~\bibnamefont
  {Hackenbroich}}, \bibinfo {author} {\bibfnamefont {P.}~\bibnamefont
  {Jacquod}},\ and\ \bibinfo {author} {\bibfnamefont {A.~D.}\ \bibnamefont
  {Stone}},\ }\bibfield  {title} {\bibinfo {title} {Semiclassical theory of the
  emission properties of wave-chaotic resonant cavities},\ }\href@noop {}
  {\bibfield  {journal} {\bibinfo  {journal} {Physical Review Letters}\
  }\textbf {\bibinfo {volume} {83}},\ \bibinfo {pages} {4991} (\bibinfo {year}
  {1999})}\BibitemShut {NoStop}%
\end{thebibliography}%
	
\end{document}